# Interpreting mode degeneracy lifting in high-index-contract optical fibers as spin-orbit interaction (SOI) or Imbert–Fedorov shift of orbital angular momentum (OAM) modes

**Liang Fang and Jian Wang***

*Wuhan National Laboratory for Optoelectronics, School of Optical and Electronic Information, Huazhong University of Science and Technology, Wuhan 430074, Hubei, China.*
*\*Corresponding author: jwang@hust.edu.cn*

**Abstract:** We present a visual interpretation of fiber mode degeneracy lifting as SOI (Imbert–Fedorov shift) of OAM modes in a ray picture, and the analytical results are verified by comparing them with eigen equation solutions.

## 1. Introduction

Spin-orbit interaction (SOI) of light is associated with spin (circularly) polarized light that affects the spatial intensity or phase distributions of light and even controls propagation paths of light [1]. It can bring about many interesting optical phenomena, such as Imbert–Fedorov (IF) or transverse spin-dependent shifts (also known as photonic spin-Hall effect) [2,3], as shown in Fig. 1(a), spin-to-orbital angular momentum (optical vortex) conversion [4], and extraordinary transverse spin at dielectric surfaces [5], etc. In optical fibers, this SOI could cause polarization plane rotation of light (optical Magnus effect) when propagating along the helically wound sing-mode fiber (SMF), which was previously used for experimental verification of Berry's phase [6].

When increasing the core size or refractive index contrast between core and cladding of fibers, the guided light meeting Maxwell's equations appears as several and even multiple eigenmodes, corresponding to the well-known few-mode or multi-mode fibers (FMF or MMF). In a straight MMF, SOI has been revealed to induce the fiber-length dependent rotation around the fiber axis of speckle patterns formed by coherent interference of many modes. In this case, the positive or negative rotation direction of speckle patterns is determined by the handedness of the spin polarization [7]. In the weakly guiding approximation (low index contrast between fiber core and cladding), as is well known, mode degeneracy makes the eigen vector modes combine into the common linearly polarized (LP) modes. These vector modes or LP modes are usually considered as modal basis of fiber-guided optical vortex that carries orbital angular momentum (OAM), and vice versa. The SOI in FMF has been investigated through indirect observation of the reflection peaks spitting of fiber Bragg gratings [8] or the length-dependent spin-controlled pattern rotation [9], both on the condition of vector mode splitting (lifting degeneracy) in high-index-contrast fibers.

Although the lifting mode degeneracy belongs to an inherent property of optical waveguides that satisfies the eigenmode equations [10], the solutions of eigenmode equations cannot provide a physical and visual insight into mode splitting of high-index-contrast fibers. Here this fiber mode degeneracy lifting is visually revealed based on SOI or IF shifts of fiber-guided OAM modes as modal basis in an approximated ray picture, as shown in Fig. 1(b).

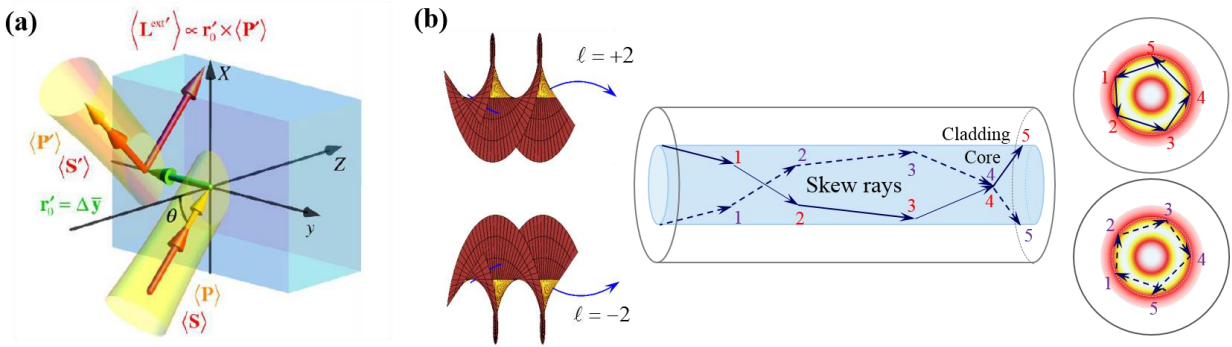

Fig. 1 **(a)** Optical SOI or spin-dependent transverse IF shift at inhomogeneous dielectric interfaces (as a result of angular momentum conservation between incident and reflected light) [2]. **(b)** Fiber-guided OAM modes can be regarded as skew ray propagation with continuously total internal reflection at the fiber core/cladding interfaces, determined by the skew Pointing vectors of OAM modes, which will also undergo SOI or IF shifts.

## 2. Analytical derivation and calculation of mode degeneracy lifting based on IF shifts of OAM modes

The skew rays of fiber-guided OAM modes ($|\ell| \neq 0$) have reverse skew directions for minus and plus OAM modes when totally reflecting at the core/cladding interface, as shown in Fig. 2(a). When undergoing IF shifts, in both cases, the OAM modes walk off for their opposite spin components, but the fundamental modes ($\ell = 0$) does not walk off due to its ray propagating on the meridian plane. The IF shift of the skew ray in Fig. 2(b) can be given by [2]

$$\delta r = -\sigma \cot\theta \cdot [1 + \text{Re}(\rho)]/k, \qquad (1)$$

where $\sigma$ denote the states of spin polarization, $\sigma = +1$ ($\sigma^+$) and $-1$ ($\sigma^-$) represents left-handed and right-handed circularly polarization, respectively. $\theta$ is the total reflection angle given by $\theta = \sin^{-1}(\beta/n_1 k)$, $n_1$ is the refractive index of fiber core, $\beta = n_{\text{eff}} k$ is the modal propagation constant of modes, $k = 2\pi/\lambda$ is the wave number of OAM modes, $n_{\text{eff}}$ is the modal effective index gotten by solving eigen equations [10]. $\rho = R_\perp / R_\parallel$, $R_\perp$ and $R_\parallel$ are complex Fresnel total reflection coefficients of parallel and vertical polarization components, respectively.

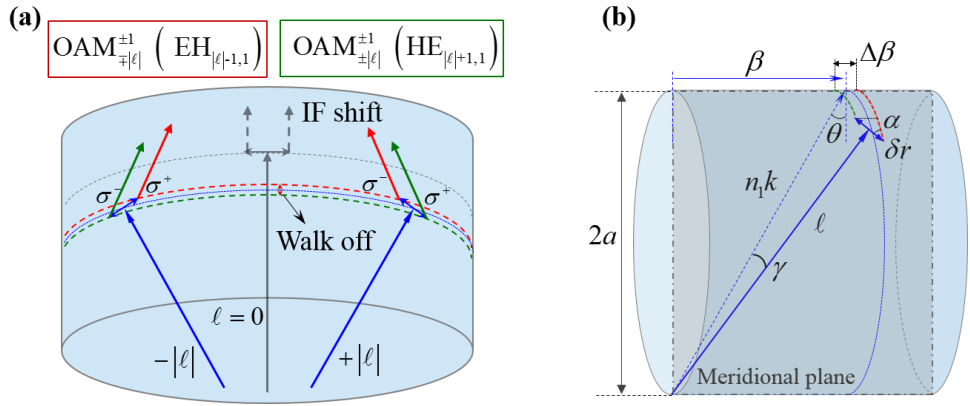

Fig 2. (a) The skew rays of Fiber-guided OAM modes ($\ell \neq 0$) undergoing SOI or IF shifts give rise to walk off between two opposite spin polarized components ($\sigma^+$ and $\sigma^-$). (b) The deflected IF shift ($\delta r$) induced by the skew angle ($\gamma$) of Pointing vectors separates the propagation constant ($\beta$) of OAM modes.

When projecting on the meridional plane in Fig. 2(b), this IF shift gives a slight propagation forward or backward for opposite spin components. Consequently, under reflection once, the propagating distance difference can be written as $\Delta z \simeq 2|\delta r| \cdot \sin\alpha$, where the deflection angle $\alpha$ is approximated as the skew angle $\gamma$ of Pointing vectors of fiber-guided OAM modes, $\alpha \approx \gamma \approx \ell/n_1 a k$, and $a$ is the radius of fiber core, similarly to the free-space OAM beams [11]. Thus the lifting degree of degeneracy based on IF shift can be approximately deduced as

$$\Delta\beta \approx \frac{\Delta z \cdot \beta}{4a\tan\theta} = \frac{\beta \cdot |\delta r| \cdot \sin\alpha}{2a\tan\theta} = \frac{|\ell|(n_1^2 k^2 - \beta^2)}{2n_1 a^2 k^2 \beta}\left[1 + \text{Re}\left(\frac{R_\perp}{R_\parallel}\right)\right]. \qquad (2)$$

It shows that the mode degeneracy lifting is mainly determined by the total reflection angle ($\theta$) and the topological charge ($\ell$) of OAM modes. The former also depends upon the index difference between core index ($n_1$) and effective index ($n_{\text{eff}}$), as shown in Fig. 2(b). Based on the total reflection condition of higher-order modes in a ray picture or eigenmode analysis in a wave picture [10], the higher index contrast between fiber core ($n_1$) and cladding ($n_2$) gives the smaller total reflection angle ($\theta$), and thus brings about the larger lifting degree of mode degeneracy.

The modal dispersions of three modal groups in two kinds of fibers are calculated by solving eigenmode equations [10], as shown in Fig. 3(a) and 3(b) where the horizontal axis is $V = ak\sqrt{n_1^2 - n_2^2}$. The modal degeneracy lifting from blue line to red and green lines corresponds to the ray splitting from blue to red and green rays in Fig. 2. To verify the deduction results in Eq. (2), the effective index differences between fiber-guided spin-orbit aligned (HE) and disaligned (EH) OAM modes are given based on the solutions of eigen equations and SOI or IF shifts as shown in Fig. 3(c). The SOI results show slight deviation relative to accurate solutions due to the ray approximation.

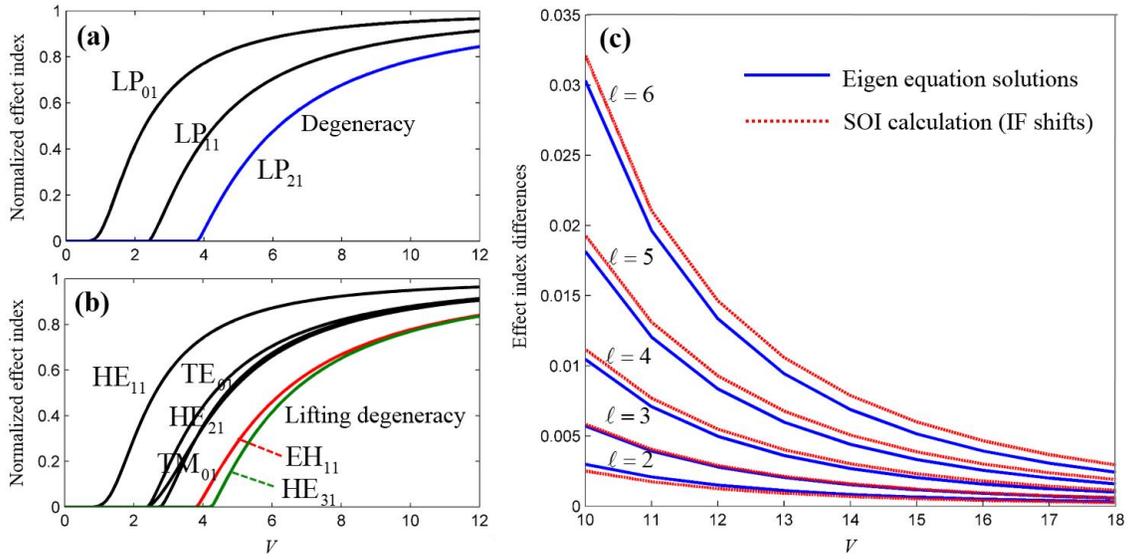

Fig. 3. Model dispersion of three modal groups in **(a)** weakly guiding optical fibers and **(b)** high-index-contrast optical fibers (cladding is air, i.e., $n_2 =1$). **(c)** Effective index difference between fiber-guided ($n_2 =1$) spin-orbit aligned (HE) and disaligned (EH) OAM modes based on the solutions of eigen equations and SOI or IF shifts here.

## 3. Discussion and conclusion

The believable results of SOI analysis and calculation in two-layer waveguides here can be easily extended to the case of three-layer OAM ring-core fibers when considering twice total reflections at both inner and outer dielectric interfaces. Even though the fiber mode degeneracy lifting is a commonplace, its underlying SOI mechanism is always ambiguous. In a ray picture, the longitudinal Goos–Hänchen shift is responsible for modifying modal propagation for LP modes. The transverse IF shift modification here provides an important supplement to reveal the mode degeneracy lifting due to SOI, as a new perspective to understand the propagation law of light in fibers.

## 4. Acknowledgements

This work was supported by the National Natural Science Foundation of China (NSFC) under grants 61905081, 61761130082, 11574001 and 11774116, .the Royal Society-Newton Advanced Fellowship, the National Program for Support of Top-notch Young Professionals.